\documentclass[preprintnumbers,eqsecnum,aps,prd,epsf,showpacs,nofootinbib]{revtex4}
\usepackage{latexsym,amsmath,amssymb}
\usepackage[dvips]{graphicx,color}
\usepackage{dcolumn,bm}    
\usepackage{subfigure}  
\newcommand{\gsim}{\mbox{\raisebox{-1.ex}{$\stackrel
      {\textstyle>}{\textstyle\sim}$}}}

\def\til#1{\tilde{#1}} 
\begin{document}
\thispagestyle{empty}
\title{Nonlinear superhorizon perturbations of non-canonical scalar
field}

\preprint{IPMU 08-0056}
\preprint{WU-AP/292/08}

\author{Yu-ichi Takamizu$^{1}$}
\email{takamizu_at_gravity.phys.waseda.ac.jp}

\author{Shinji Mukohyama$^{2}$}
\email{shinji.mukohyama_at_ipmu.jp}
\affiliation{
\\
$^{1}$ Department of Physics, Waseda University,
Okubo 3-4-1, Shinjuku, Tokyo 169-8555, Japan
\\
$^{2}$ Institute for the Physics and Mathematics of the Universe (IPMU),
The University of Tokyo, 5-1-5 Kashiwanoha, Kashiwa, 
Chiba 277-8582, Japan
}
\date{\today}

\begin{abstract}
 We develop a theory of non-linear cosmological perturbations at
 superhorizon scales for a scalar field with a Lagrangian of the form
 $P(X,\phi)$, where $X=-\partial^{\mu}\phi\partial_{\mu}\phi$ and $\phi$
 is the scalar field. We employ the ADM formalism and the spatial
 gradient expansion approach to obtain general solutions valid up to
 the second order in the gradient expansion. This formulation can be
 applied to, for example, DBI inflation models to investigate
 superhorizon evolution of non-Gaussianities. With slight modification,
 we also obtain general solutions valid up to the same order for a
 perfect fluid with a general equation of state $P=P(\rho)$. 
\end{abstract}
\maketitle

\section{Introduction}
\label{sec:intro}

Generation of primordial fluctuations during inflation is one
of the most interesting predictions of quantum field theory. Indeed,
those quantum fluctuations are considered as seeds of the large scale
structure of the present universe, and this picture has been accepted by
many researchers as a standard scenario. The cosmic microwave background
(CMB) temperature anisotropy found by COBE~\cite{Smoot:1992td} was
perfectly consistent with the predictions of the linear theory of
cosmological
perturbations~\cite{Bardeen:1980kt,Kodama:1985bj,Mukhanov:1990me}. In
particular, the primordial fluctuations are nearly scale invariant and
consistent with Gaussian statistics~\cite{Spergel:2006hy}.

The recent more accurate observation by WMAP, however, has revealed 
deviation from exact scale invariance, with a slight red
tilt~\cite{Komatsu:2008hk}. Moreover, there is a good possibility that
deviation from Gaussianity can be detected by the future experiments
such as PLANCK~\cite{:2006uk}. With those current and future precision
observations, deviation from the exact scale invariance and Gaussianity
can be a powerful tool to discriminate many possible inflationary
models. Especially, non-Gaussianity of primordial fluctuations has
recently been a focus of much attention by many 
authors~\cite{Bartolo:2004if,Maldacena:2002vr,Seery:2005wm,Malik:2003mv,Rigopoulos:2003ak,Lyth:2004gb,Langlois:2005ii,Lyth:2005fi,Komatsu:2001rj,Vernizzi:2006ve,Rigopoulos:2005xx,Battefeld:2006sz,Yokoyama:2007uu,Sasaki:2008uc,Arroja:2008yy,Langlois:2008qf,Tanaka:2006zp,Tanaka:2007gh}.

In order to parameterize the amount of non-Gaussianity of primordial
perturbations, commonly used is the non-linear parameter $f_{NL}$. This is
related to the bi-spectrum of the curvature perturbation, and is defined
as~\cite{Komatsu:2001rj}
\begin{align}
\zeta({\bf x})=\zeta_{G}({\bf x})-{3\over 5}f_{NL} \zeta^2_G({\bf x})\,,
\end{align}
where $\zeta_G$ is the curvature perturbation on a uniform density
hypersurface and satisfies linear Gaussian statistics. 
On the observational side, current bounds on the parameter $f_{NL}$ by
WMAP five years~\cite{Komatsu:2008hk} are $-9<f_{NL}<111$ for the local
form of bi-spectrum and $ -151<f_{NL}<253$ for the equilateral form. By future
experiments such as PLANCK \cite{:2006uk}, it is expected that
non-Gaussianity of the level $|f_{NL}|\gsim 5$ can be
detected~\cite{Komatsu:2001rj}.

On the theoretical side, although the non-Gaussianity from the standard
single-field slow-roll inflation is suppressed by slow-roll
parameters~\cite{Maldacena:2002vr,Lyth:2005fi} and is too small to be 
detected in near-future experiments, many new types of inflationary
models predicting large non-Gaussianity ($f_{NL}$ greater than unity)
have been proposed. There are at least two known mechanisms to generate
large non-Gaussianity: isocurvature perturbations and non-canonical
kinetic terms. Intriguingly, bi-spectrums for these two mechanism
typically have different shapes: the so called local type for
non-Gaussianity from isocurvature perturbations and the equilateral type
for non-Gaussianity from non-canonical kinetic terms.

As for the former case, a typical example is the curvaton
scenario~\cite{Moroi:2001ct,Lyth:2001nq,Malik:2006pm,Sasaki:2006kq},
where a light scalar field, called curvaton, is responsible for
isocurvature perturbations during inflation. Another example is
multi-field inflation models, where generation of isocurvature 
perturbations is generally expected because of the existence of more
than one fields in the inflaton
sector~\cite{Rigopoulos:2005xx,Vernizzi:2006ve,Battefeld:2006sz,Yokoyama:2007uu}.
In these examples, isocurvature perturbations generated during inflation
are later converted to curvature perturbations, and this process is
important for large non-Gaussianity. Now let us consider the later case,
where non-canonical kinetic terms are responsible for large
non-Gaussianity. Examples of this type include
k-inflation~\cite{ArmendarizPicon:1999rj,Garriga:1999vw}, ghost 
inflation~\cite{ArkaniHamed:2003uz,Senatore:2004rj} and DBI
inflation~\cite{Silverstein:2003hf,Alishahiha:2004eh}. 
In fact, Weinberg \cite{Weinberg:2008hq} has recently shown that 
the leading corrections to the Gaussian correlations of curvature 
perturbation are solely of the k-inflation type (including DBI
inflation), except for ghost inflation. In k-inflation and DBI
inflation, large non-Gaussianity is expected when the non-linear nature
of the non-canonical kinetic action becomes significant. This happens
when the sound speed of perturbations is sufficiently smaller than
unity \cite{Chen:2006nt}.

On the other hand, in ghost inflation large non-Gaussianity, such as
$|f_{NL}|\simeq 80$, is always expected {\it unless} fine-tuned. The
reason is that non-linear terms in the low energy effective Lagrangian
is suppressed only by fractional (not integer) powers of energy
scales. To be more precise, the leading non-linear term is suppressed
only by $(H/M)^{1/4}$, where $H$ is the Hubble expansion rate during 
inflation and $M$ is the cutoff scale of the low energy effective
theory. This is confirmed by a simple scaling analysis and, thus,
should be robust. After using the COBE normalization
$\delta\rho/\rho\simeq 10^{-5}$, which implies $H/M\simeq 10^{-4}$,
this fact leads to the prediction of large
non-Gaussianity~\cite{ArkaniHamed:2003uz} if the dimensionless coupling 
constant for the leading nonlinear term is set to be order unity,
i.e. if we do {\it not} fine-tune the theory.

For these reasons, non-Gaussianity is one of the most powerful tools to
distinguish models of inflation with combination of the future 
observations. Thus, to quantify the non-Gaussianity and clarify its
observational signature, it is important to develop a theory that can
deal with nonlinear cosmological perturbations. There are couple of
methods to tackle this problem. One is a second-order perturbation 
theory~\cite{Maldacena:2002vr,Acquaviva:2002ud}.  Another is based on 
spatial gradient expansion~\cite{Lyth:2004gb,Salopek:1990jq}. While the
former mainly deals with primordial perturbations up to around the
horizon exit, the later deals with classical evolution after horizon
exist. Thus it is important to develop both methods and to use them
complementarily.

Closely related to the gradient expansion method, cosmological
perturbations on superhorizon scales have been studied extensively in
the so-called separate universe approach or $\delta N$
formalism~\cite{Sasaki:1998ug,Wands:2000dp}. Actually, these approaches
are essentially the leading order approximation to the gradient
expansion~\cite{Rigopoulos:2003ak,Lyth:2004gb}. Including these, many of 
the previous studies were confined to the leading order approximation to
the gradient expansion. However, higher order corrections to the leading
order results can be important to get more detailed information about
non-Gaussianity. One good example is the case studied by Leach et al
\cite{Leach:2001zf}. They considered linear perturbations
in single-field inflation models and supposed that there is a stage at
which slow-roll conditions are violated. It has been then shown that,
due to the decaying mode, the $O(\epsilon^2)$ corrections in spatial
derivative expansion do affect the evolution of curvature perturbations
on superhorizon scales~\cite{Leach:2001zf}. A similar situation for DBI 
inflation was considered by Jain et al~\cite{Jain:2007au}. In these
situations, it is expected that non-Gaussianities should also be
affected by the order $O(\epsilon^2)$ effects. However, the linear
perturbation theory is not capable for calculation of
non-Gaussianity. Thus, it is necessary to develop nonlinear theory of
cosmological perturbations valid up to $O(\epsilon^2)$ in the spatial
gradient expansion.

Gradient expansion formalism has been developed and used by many
authors~\cite{Lyth:2004gb,Salopek:1990jq,Sasaki:1998ug,Wands:2000dp,Tanaka:2006zp,Tanaka:2007gh}. 
Formulation valid up to $O(\epsilon^2)$ was developed, for example, by
Tanaka and Sasaki for a universe dominated by a perfect fluid with a
specific equation of state $P/\rho=const$~\cite{Tanaka:2006zp} and that
dominated by a canonical scalar field~\cite{Tanaka:2007gh}. However, as
far as the authors know, those works have not extended to a perfect
fluid with general equation of state $P=P(\rho)$ nor to a scalar field
with non-canonical kinetic action, which is essential for the second
type of mechanism of generating non-Gaussianity.

The purpose of this paper is to fill this gap. Namely, we shall develop
a theory of nonlinear superhorizon perturbations valid up to the order
$O(\epsilon^2)$ for a scalar field with non-canonical kinetic action and
a perfect fluid with general equation of state.

This paper is organized as follows. In Sec.~\ref{sec:scalarfield}, we introduce a non-canonical scalar field as our model 
and express it in a perfect fluid form. In 
Sec.~\ref{sec:formulation}, we shall develop a theory of nonlinear 
cosmological perturbations on superhorizon scales and explain our 
formulation in 
details. The following Sec.~\ref{sec:solution} 
is devoted to some details of 
obtaining a general solution.  We then study some specific examples 
in Sec.~\ref{sec:examples}. Sec.~\ref{sec:summary} is devoted to a
summary of this paper and discussion. In Appendix, we give our result in 
a perfect fluid system. 

\section{Scalar field in a perfect fluid form}
\label{sec:scalarfield}

Throughout this paper we consider a minimally-coupled scalar field
described by an action of the form
%
\begin{equation}
 I = \int d^4x\sqrt{-g}P(X,\phi), 
\end{equation}
where $X=-g^{\mu\nu}\partial_{\mu}\phi\partial_{\nu}\phi$, and suppose
that $-g^{\mu\nu}\partial_{\nu}\phi$ is timelike and future-directed. The 
equation of motion for $\phi$ is
%
\begin{equation}
 \frac{2}{\sqrt{-g}}\partial_{\mu}\left(\sqrt{-g}P_X\partial^{\mu}\phi\right)
  + P_{\phi} = 0,
  \label{eqn:EOM-phi}
\end{equation}
where the subscripts $X$ and $\phi$ represent derivative with respect to
$X$ and $\phi$, respectively. The stress energy tensor of the scalar
field is shown to be a perfect fluid form:
%
\begin{equation}
 T_{\mu\nu} = 2P_X\partial_{\mu}\phi\partial_{\nu}\phi + Pg_{\mu\nu}
  = (\rho+P)u_{\mu}u_{\nu} + Pg_{\mu\nu},
  \label{eqn:Tmunu-phi}
\end{equation}
where 
%
\begin{equation}
 \rho(X,\phi) = 2P_X X - P, \quad
  u_{\mu} = -\frac{\partial_{\mu}\phi}{\sqrt{X}}.
 \end{equation}
Note that $u^{\mu}u_{\mu}=-1$. As far as $\partial_{\mu}\phi\ne 0$, the
equation of motion (\ref{eqn:EOM-phi}) is equivalent to the conservation
equation $\nabla_{\mu}T^{\mu}_{\ \nu}=0$.

The following relation among first-order variations of $P$, $\rho$ and
$\phi$ will be useful in the analysis below. 
%
\begin{equation}
 \delta P = c_s^2\delta\rho + \rho\Gamma\delta\phi,
  \label{eqn:deltaP}
 \end{equation}
where
%
\begin{equation}
 c_s^2 = \frac{P_X}{2P_{XX}X+P_X}, \quad
  \Gamma = \frac{1}{\rho}\left(P_{\phi}-c_s^2\rho_{\phi}\right). 
\end{equation}
Note that $c_s$ is the speed of sound for the gauge invariant scalar
perturbation in the linear theory~\cite{Garriga:1999vw}.

\section{Formalism}
\label{sec:formulation}

In this section we shall develop a theory of nonlinear cosmological
perturbations on superhorizon scales. For this purpose we employ the ADM
formalism and the gradient expansion in the uniform Hubble slicing.

\subsection{ADM decomposition}

In the ($3+1$)-decomposition, the metric is expressed as 
%
\begin{equation}
 ds^2 = g_{\mu\nu}dx^{\mu}dx^{\nu}
  = - \alpha^2 dt^2 + \gamma_{ij}(dx^i+\beta^idt)(dx^j+\beta^jdt), 
\end{equation}
where $\alpha$ is the lapse function, $\beta^i$ is the shift vector and
Latin indices run over $1, 2, 3$. Since $\alpha$ and $\beta^i$ represent
gauge degrees of freedom for diffeomorphism and appear as Lagrange
multipliers in the action, the corresponding equations of motion leads
to constraint equations. Contrary to $\alpha$ and $\beta$, components of
the spatial metric $\gamma_{ij}$ are dynamical variables (subject to the
constraint equations)  and the corresponding equations of motion  are
called dynamical equations. In what follows we shall express the
dynamical equations as a set of first-order differential equations with
respect to the time $t$. For this purpose we introduce the extrinsic
curvature $K_{ij}$ defined by  
%
\begin{equation}
 K_{ij} =
  -\frac{1}{2\alpha}\left(\partial_t\gamma_{ij}-D_i\beta_j-D_j\beta_i\right),
  \label{eqn:def-K}
\end{equation}
where $D$ is the covariant derivative compatible with the spatial metric
$\gamma_{ij}$. For the stress-energy tensor in the perfect fluid 
form (\ref{eqn:Tmunu-phi}), we define the $3$-vector $v^i$ as
$v^i\equiv u^i/u^0$. Hereafter, we shall use $\gamma_{ij}$ and its
inverse $\gamma^{ij}$ to raise and lower indices of $K$, $D$, $v$,
$\beta$. Then we can express $u^{\mu}$ and $u_{\mu}$ in terms of
$\alpha$, $\beta^i$ and $v^i$: 
%
\begin{eqnarray}
 u^0 & = & \left[\alpha^2-(v_k+\beta_k)(v^k+\beta^k)\right]^{-1/2}, 
  \nonumber\\
 u^i & = & u^0v^i, \nonumber\\
 u_0 & = & -u^0\left[\alpha^2-\beta_k(v^k+\beta^k)\right], \nonumber\\
 u_i & = & u^0(v_i+\beta_i).
  \label{eqn:u-v}
\end{eqnarray}
The conservation equation $\nabla_{\mu}T^{\mu}_{\ \nu}=0$ is
%
\begin{eqnarray}
 u^{\mu}\partial_{\mu}\rho + \frac{\rho+P}{\alpha\sqrt{\det\gamma}}
  \partial_{\mu}\left(\alpha\sqrt{\det\gamma}u^{\mu}\right) & = & 0,
  \nonumber\\
 \frac{1}{\sqrt{\det\gamma}}\partial_t\left[\sqrt{\det\gamma}(\rho+P)wu_i\right]
  + D_j\left[(\rho+P)wv^ju_i\right] 
  & = & - \alpha\partial_iP
  - (\rho+P)w\left[w\partial_i\alpha-u_jD_i\beta^j\right],
\end{eqnarray}
where $w\equiv \alpha u^0$. All independent components of the stress
energy tensor are conveniently expressed as
%
\begin{eqnarray}
 E & \equiv & T_{\mu\nu}n^{\mu}n^{\nu} = (\rho+P)w^2-P, \nonumber\\
 J_i & \equiv & -T_{\mu i}n^{\mu} = (\rho+P)wu_i, \nonumber\\
 S_{ij} & \equiv & T_{ij} = (\rho+P)u_iu_j+P\gamma_{ij},
\end{eqnarray}
where $n^{\mu}$ is the unit vector normal to the constant $t$ surfaces
and is given by 
%
\begin{equation}
 n_{\mu}dx^{\mu} = -\alpha dt, \quad
  n^{\mu}\partial_{\mu} = \frac{1}{\alpha}(\partial_t-\beta^i\partial_i).
\end{equation}

The Hamiltonian constraint, corresponding to the equation of motion for
$\alpha$, is  
%
\begin{equation}
 R + K^2 - K_{ij}K^{ij} = 2\kappa^2 E, 
   \label{eqn:Hamiltonian-const}
\end{equation}
where $\kappa^2=8\pi G_N$, $R\equiv R[\gamma]$ is the Ricci scalar of
the spatial metric $\gamma$ and $K\equiv K^i_{\ i}$. The momentum
constraint, corresponding to the equation of motion for $\beta^i$, is 
%
\begin{equation}
 -\partial_iK + D_jK^j_{\ i} = \kappa^2 J_i.
  \label{eqn:Momentum-const}
\end{equation}
The dynamical equations, decomposed into the trace part and the
traceless part, are  
%
\begin{eqnarray}
 \partial_{\perp}K - K_{ij}K^{ij}+ \frac{D^2\alpha}{\alpha}
  & = & \frac{\kappa^2}{2}\left(S^k_{\ k}+E\right), \nonumber\\
 R_{ij} - \nabla_{\perp}K_{ij}+ KK_{ij} - \frac{D_iD_j\alpha}{\alpha}
  - \frac{\gamma_{ij}}{3}
  \left[R-\partial_{\perp}K+K^2-\frac{D^2\alpha}{\alpha}\right]
 & = & \kappa^2\left(S_{ij}- \frac{1}{3}S^k_{\ k}\gamma_{ij}\right),
 \label{eqn:dynamica-eqs}
\end{eqnarray}
where $\partial_{\perp}\equiv n^{\mu}\partial_{\mu}$, 
$\nabla_{\perp}\equiv n^{\mu}\nabla_{\mu}$, 
$D^2\equiv\gamma^{ij}D_iD_j$ and $R_{ij}\equiv R_{ij}[\gamma]$ is the
Ricci tensor of the spatial metric $\gamma$.

In addition to the standard ADM decomposition briefly reviewed above, we
further decompose the spatial metric and the extrinsic curvature as
%
\begin{eqnarray}
 \gamma_{ij} & = & a^2\psi^4\tilde{\gamma}_{ij}, 
  \nonumber\\
 K_{ij} & = & 
  a^2\psi^4\left(\frac{1}{3}K\tilde{\gamma}_{ij}
	    +\tilde{A}_{ij}\right),
	     \label{eqn:decompose-metric}
\end{eqnarray}
where $a(t)$ is the scale factor of a fiducial Friedmann background
(specified later) and the determinant of $\tilde{\gamma}_{ij}$ is
constrained to be unity: $\det\tilde{\gamma}_{ij}=1$. The first-order 
equations for the spatial metric ($\psi$, $\tilde{\gamma}_{ij}$) are
deduced from the definition of the extrinsic curvature (\ref{eqn:def-K})
as 
%
\begin{eqnarray}
 \frac{\partial_{\perp}\psi}{\psi} + \frac{\partial_t a}{2\alpha a}
  & = &
  \frac{1}{6}\left(-K+\frac{\partial_i\beta^i}{\alpha}\right),
  \label{eqn:dpsi}\\
 \partial_{\perp}\tilde{\gamma}_{ij} & = & -2\tilde{A}_{ij}
  + \frac{1}{\alpha}
  \left(\tilde{\gamma}_{ik}\partial_j\beta^k
   +\tilde{\gamma}_{jk}\partial_i\beta^k
   -\frac{2}{3}\tilde{\gamma}_{ij}\partial_k\beta^k\right).
  \label{eqn:dgammatilde}
\end{eqnarray}
The first-order equations for the extrinsic curvature ($K$,
$\tilde{A}_{ij}$) are obtained from the dynamical equations
(\ref{eqn:dynamica-eqs}) as
%
\begin{eqnarray}
 \partial_{\perp}K & = &  
  \frac{K^2}{3} + \tilde{A}^{ij}\tilde{A}_{ij} - \frac{D^2\alpha}{\alpha} 
  + \frac{\kappa^2}{2}\left(S^k_{\ k}+E\right), 
  \label{eqn:dynam-K-K}\\
  \partial_{\perp}\tilde{A}_{ij} & = & 
   \left(K\tilde{A}_{ij}-2\tilde{A}_i^{\ k}\tilde{A}_{kj}\right)
   + \frac{1}{\alpha}\left(\tilde{A}_{ik}\partial_j\beta^k
   +\tilde{A}_{jk}\partial_i\beta^k
   -\frac{2}{3}\tilde{A}_{ij}\partial_k\beta^k\right) \nonumber\\
 & & 
   +\frac{1}{a^2\psi^4}
   \left[\left(R_{ij}-\frac{R}{3}\gamma_{ij}\right)
    -\frac{1}{\alpha}\left(D_iD_j\alpha-\frac{D^2\alpha}{3}\gamma_{ij}\right)
    - \kappa^2\left(S_{ij}-\frac{S^k_{\ k}}{3}\gamma_{ij}\right)
       \right],
        \label{eqn:dynam-K-tildeA}
\end{eqnarray}
where $\tilde{A}_i^{\ k}=\tilde{\gamma}^{jk}\tilde{A}_{ij}$ and
$\tilde{A}^{jk}=\tilde{\gamma}^{ij}\tilde{A}_i^{\ k}$. The Hamiltonian
and momentum constraints are, respectively, 
%
\begin{eqnarray}
 R + \frac{2}{3}K^2 - \tilde{A}^{ij}\tilde{A}_{ij} & = & 2\kappa^2E, 
  \label{eqn:Hamiltonian-constraint} \\
 -\frac{2}{3}\partial_iK + D_j\tilde{A}_i^{\ j}
  & = & \kappa^2J_i. 
  \label{eqn:momentum-constraint}
\end{eqnarray}
The conservation equation is
%
\begin{eqnarray}
 \left(\partial_t+v^i\partial_i\right)\rho
  + \frac{\rho+P}{(a\psi^2)^3w}
  \left\{\partial_t\left[(a\psi^2)^3w\right]
   +\partial_i\left[(a\psi^2)^3wv^i\right]\right\} & = & 0, 
  \label{eqn:conservation-t} \\
 \frac{1}{(a\psi^2)^3}\partial_t\left[(a\psi^2)^3(\rho+P)wu_i\right]
  + D_j\left[(\rho+P)wv^ju_i\right] + \partial_iP
  + (\rho+P)(w^2\partial_i\alpha-wu_jD_i\beta^j) & = & 0.
  \label{eqn:conservation-i} 
\end{eqnarray}

Throughout this paper we adopt the uniform Hubble slicing 
%
\begin{equation}
 K = -3H(t), \quad H(t)\equiv \frac{\partial_t a}{a}. 
 \label{eqn:uniform-Hubble}
\end{equation}
Substituted into (\ref{eqn:dpsi}), this implies that 
%
\begin{equation}
 \chi \ (\equiv \alpha - 1) = \frac{2\partial_t\psi}{H\psi} -
  \frac{D_i\beta^i}{3H}. \label{eqn:uniform-H}
\end{equation}

\subsection{Gradient expansion: basic assumptions and order estimates}

In the gradient expansion approach we introduce a flat FRW universe
($a(t)$, $\phi_0(t)$) as a background  and suppose that the
characteristic length scale $L$ of perturbations is longer than the
Hubble length scale $1/H$ of the background, i.e.  $HL\gg 1$. Therefore,
we consider $\epsilon\equiv 1/(HL)$ as a small parameter and
systematically expand our equations by $\epsilon$, considering a spatial
derivative acted on perturbations is of order $O(\epsilon)$. 

The background flat FRW universe ($a(t)$, $\phi_0(t)$) satisfies the
Friedmann equation and the equation of motion
%
\begin{equation}
 H^2 = \frac{\kappa^2}{3}\rho_0, \quad 
 \frac{2}{a^3}\partial_t\left(a^3P_{0X}\partial_t\phi_0\right) -
 P_{0\phi} = 0,
 \label{eqn:background-eq}
\end{equation}
where $\rho_0\equiv \rho(X_0,\phi_0)$, 
$P_{0X}\equiv P_X(X_0,\phi_0)$, 
$P_{0\phi}\equiv P_{\phi}(X_0,\phi_0)$,  and
$X_0\equiv(\partial_t\phi_0)^2$.

Since the FRW background is recovered in the limit $\epsilon\to 0$, we
naturally have the estimates
%
\begin{equation}
 v^i=O(\epsilon), \quad \beta^i=O(\epsilon), 
  \label{eqn:assumption-vbeta}
\end{equation}
and $\partial_t\tilde{\gamma}_{ij}=O(\epsilon)$. Actually, following the
arguments in refs.~\cite{Tanaka:2006zp,Tanaka:2007gh},  we assume a
stronger condition
%
\begin{equation}
 \partial_t\tilde{\gamma}_{ij} = O(\epsilon^2). 
 \label{eqn:assumption-gamma}
\end{equation}
This assumption significantly simplifies our analysis and, we believe,
still allows many useful applications of the formalism. On the other
hand, we consider $\psi$ and $\tilde{\gamma}_{ij}$ (without derivatives
acted on them) as quantities of order $O(1)$. 

We can estimate orders of magnitude of various quantities by using the
above assumption and the basic equations. First, (\ref{eqn:dgammatilde})
implies that 
%
\begin{equation}
 \tilde{A}_{ij}=O(\epsilon^2). \label{eqn:Atilde-epsilon2}
\end{equation}
Substituting (\ref{eqn:Atilde-epsilon2}) into
(\ref{eqn:momentum-constraint}) we obtain $J_i=O(\epsilon^3)$, or 
%
\begin{equation}
 v_i+\beta_i=O(\epsilon^3). \label{eqn:vbeta-epsilon3}
\end{equation}
For the scalar field system, this is expressed as
$\partial_i\pi=O(\epsilon^3)$, where $\pi\equiv \phi-\phi_0$. By
absorbing a homogeneous part of $\pi$ into $\phi_0$ (and modifying
$a(t)$ accordingly), we obtain 
%
\begin{equation}
 \pi = O(\epsilon^2). 
\end{equation}
Combining (\ref{eqn:vbeta-epsilon3}) with the first equation in
(\ref{eqn:u-v}), we obtain $u^0=1/\alpha+O(\epsilon^6)$, or 
%
\begin{equation}
 w \ (\equiv \alpha u^0) = 1 + O(\epsilon^6). 
\end{equation}
This implies that $E=\rho+O(\epsilon^6)$. Thus, from
(\ref{eqn:Hamiltonian-constraint}) we obtain
%
\begin{equation}
 \delta \ \left(\equiv \frac{\rho-\rho_0}{\rho_0}\right) =
  O(\epsilon^2), 
\end{equation}
and (\ref{eqn:deltaP}) implies that
%
\begin{equation}
 p \ (\equiv P-P_0) = O(\epsilon^2). 
\end{equation}
Finally, (\ref{eqn:conservation-t}) implies that
%
\begin{equation}
 \partial_t\psi = O(\epsilon^2),
\end{equation}
and thus we obtain
%
\begin{equation}
 \chi = O(\epsilon^2)
\end{equation}
from the uniform Hubble slicing condition (\ref{eqn:uniform-H}). 

In summary, we have the following estimates (including assumptions):
%
\begin{eqnarray}
 & & 
  \psi = O(1), \quad \tilde{\gamma}_{ij} = O(1), \quad v^i=O(\epsilon),
  \quad \beta^i=O(\epsilon), \nonumber\\
 & & 
  \chi = O(\epsilon^2), \quad \tilde{A}_{ij} = O(\epsilon^2), 
  \quad \delta =O(\epsilon^2), \quad \pi=O(\epsilon^2),  
  \quad p = O(\epsilon^2), \nonumber\\
 & & 
  \partial_t\tilde{\gamma}_{ij}=O(\epsilon^2), \quad
  \partial_t\psi=O(\epsilon^2), \quad
  v^i+\beta^i = O(\epsilon^3),  \quad w = 1 + O(\epsilon^6). 
  \label{eqn:order-of-magnitude}
\end{eqnarray}

\subsection{Leading order equations}

Substituting the order of magnitude shown in
(\ref{eqn:order-of-magnitude}) into the conservation equations
(\ref{eqn:conservation-t}) and (\ref{eqn:conservation-i}), we find 
%
\begin{eqnarray}
 \rho_0\partial_t \delta
  +(\rho_0+P_0)\left(6\frac{\partial_t\psi}{\psi}+D_iv^i\right)
  + 3H(p-P_0\delta) & = & O(\epsilon^4), 
  \label{eqn:hydro1}\\
 \frac{1}{a^3}\partial_t\left[a^3(\rho_0+P_0)u_i\right]
  + \partial_i \left[p + (\rho_0+P_0)\chi\right] & = & O(\epsilon^5).
  \label{eqn:hydro2}
\end{eqnarray}
The Hamiltonian and momentum constraint equations give  
%
\begin{eqnarray}
 8\frac{\tilde{D}^2\psi}{\psi} & = & 
  \tilde{R} - 2\kappa^2 a^2\psi^4\rho_0\delta + O(\epsilon^4),
  \label{eqn:Hamiltonian}\\
 \tilde{D}_j\left(\psi^6\tilde{A}_{i}^{\ j}\right)
  & = & 
  \kappa^2(\rho_0+P_0)\psi^6u_i + O(\epsilon^5), 
  \label{eqn:Momentum}
\end{eqnarray}
where $\tilde{R}\equiv R[\tilde{\gamma}]$ is the Ricci scalar of the
normalized spatial metric $\tilde{\gamma}_{ij}$, $\tilde{D}$ is the
covariant derivative compatible with $\tilde{\gamma}_{ij}$, 
$\tilde{D}^2\equiv\tilde{\gamma}^{ij}\tilde{D}_i\tilde{D}_j$, and
$\tilde{\gamma}^{ij}$ is the inverse matrix of
$\tilde{\gamma}_{ij}$. The evolution equations for the spatial metric 
give 
%
\begin{eqnarray}
 6\frac{\partial_t\psi}{\psi} & = & 3H\chi + D_i\beta^i + O(\epsilon^4), 
  \label{eqn:evol-gamma1}\\
 (\partial_t-\beta^k\partial_k)\tilde{\gamma}_{ij} & = & 
  -2\tilde{A}_{ij}+\tilde{\gamma}_{ik}\partial_j\beta^k
  +\tilde{\gamma}_{jk}\partial_i\beta^k
  -\frac{2}{3}\tilde{\gamma}_{ij}\partial_k \beta^k
  +O(\epsilon^4),
  \label{eqn:evol-gamma2}
\end{eqnarray}
while the evolution equations for the extrinsic curvature give 
%
\begin{eqnarray}
 \partial_t\til{A}_{ij}+3H\tilde{A}_{ij} & = &
  \frac{1}{a^2\psi^4}\left(R_{ij}-\frac{1}{3}R\gamma_{ij}\right)
  +O(\epsilon^4),  \label{eqn:-evol-K1}\\
 p + (\rho_0+P_0)\chi & = & - \frac{1}{3}\rho_0\delta + O(\epsilon^4),
  \label{eqn:evol-K2}
\end{eqnarray}

By using (\ref{eqn:evol-gamma1}), (\ref{eqn:evol-K2}) and the background
conservation equation $\partial_t\rho_0+3H(\rho_0+P_0)=0$, a single
equation for $\delta$ is easily obtained from (\ref{eqn:hydro1}),  
%
\begin{equation}
 \partial_t(a^2\rho_0\delta)  = O(\epsilon^4).
  \label{eqn:eq-for-delta}
\end{equation}
Using (\ref{eqn:evol-K2}) again, (\ref{eqn:hydro2}) is simplified to 
%
\begin{equation}
 \frac{1}{a^3}\partial_t\left[a^3(\rho_0+P_0)u_i\right] =  
  \frac{1}{3}\rho_0\partial_i \delta + O(\epsilon^5).
  \label{eqn:eq-for-ui}
\end{equation}
It is intriguing to note that we have not yet specified the form of
$p$.  Therefore, eqs.~(\ref{eqn:hydro1})-(\ref{eqn:eq-for-ui}) hold as
far as the stress-energy tensor is of the perfect fluid form and
$p=O(\epsilon^2)$, regardless of whether the stress-energy tensor is
provided by a scalar field, radiation, dust, or any other sources.

The form of $p$ for the scalar field system is specified by the relation
(\ref{eqn:deltaP}) as 
%
\begin{equation}
 p = \rho_0(c_{s0}^2\delta + \Gamma_0\pi) + O(\epsilon^4),
  \label{eqn:p-delta-pi}
\end{equation}
where $c_{s0}^2=P_{0X}/(2P_{0XX}X_0+P_{0X})$ and  
$\Gamma_0=(P_{0\phi}-c_{s0}^2\rho_{0\phi})/\rho_0$. 
We can obtain another equation relating $p$ and $\pi$, by expanding $p$
as $p=P_{0X}(X-X_0)+P_{0\phi}\pi+O(\epsilon^4)$, where
$X-X_0=2(\partial_t\phi_0\partial_t\pi-\chi X_0)+O(\epsilon^4)$. 
Actually, this equation can be interpreted as a first-order equation for
$\pi$. Using (\ref{eqn:evol-K2}), $\rho_0+P_0=2P_{0X}X_0$ and the
background equation of motion (\ref{eqn:background-eq}), we can rewrite 
this equation for $\pi$ as  
%
\begin{equation}
 \frac{1}{a^3}\partial_t\left[\frac{(\rho_0+P_0)a^3}{\partial_t\phi_0}\pi\right]
  = - \frac{1}{3}\rho_0\delta + O(\epsilon^4). 
  \label{eqn:eq-pi}
\end{equation}

\section{General solution}
\label{sec:solution}

Having written down all relevant equations up to the order
$O(\epsilon^2)$ in the gradient expansion, we now seek a general
solution.  

\subsection{Leading order solutions}

First, $\psi=O(1)$ and $\partial_t\psi=O(\epsilon^2)$ imply that 
%
\begin{equation}
 \psi = L^{(0)}(x^k) + O(\epsilon^2),
  \label{eqn:psi-leading}
\end{equation}
where $L^{(0)}(x^k)$ is an arbitrary function of the spatial coordinates
$\{x^k\}$ ($k=1,2,3$). Hereafter, the superscript $(n)$ indicates that
the corresponding quantity is of order $O(\epsilon^n)$. Similarly, 
$\tilde{\gamma}_{ij}=O(1)$ and
$\partial_t\tilde{\gamma}_{ij}=O(\epsilon^2)$ imply that
%
\begin{equation}
 \tilde{\gamma}_{ij} = f^{(0)}_{ij}(x^k) + O(\epsilon^2),
  \label{eqn:gammatilde-leading}
\end{equation}
where $f^{(0)}_{ij}(x^k)$ is a ($3\times 3$)-matrix with unit
determinant whose components depend only on the spatial
coordinates. With these expressions, the right hand side of
(\ref{eqn:-evol-K1}) is calculated as 
%
\begin{equation}
 \frac{1}{a^2\psi^4}\left(R_{ij}-\frac{1}{3}R\gamma_{ij}\right)
  = \frac{1}{a^2}F^{(2)}_{ij}(x^k) + O(\epsilon^4),
\end{equation}
where
%
\begin{eqnarray}
 F^{(2)}_{ij}(x^k) & \equiv & \frac{1}{\left(L^{(0)}\right)^4}
  \left[ \left(\tilde{R}^{(0)}_{ij}-\frac{1}{3}\tilde{R}^{(0)}f^{(0)}_{ij}\right)
   + 2\left(2\partial_i\ln L^{(0)}\partial_j\ln L^{(0)}
   - \tilde{D}^{(0)}_i\tilde{D}^{(0)}_j\ln L^{(0)}\right)\right.\nonumber\\
 & & \left.
   - \frac{2}{3}f_{(0)}^{kl}
   \left(2\partial_k\ln L^{(0)}\partial_l\ln L^{(0)}
   - \tilde{D}^{(0)}_k\tilde{D}^{(0)}_l\ln L^{(0)} \right) f^{(0)}_{ij}
  \right],
 \label{eqn:def-F2ij}
\end{eqnarray}
$f_{(0)}^{kl}$ is the inverse matrix of $f^{(0)}_{ij}$,
$\tilde{R}^{(0)}_{ij}=R_{ij}[f^{(0)}]$ and $\tilde{R}^{(0)}=R[f^{(0)}]$
are Ricci tensor and Ricci scalar of the $0$th-order spatial metric
$f^{(0)}_{ij}$, and $\tilde{D}^{(0)}$ is the covariant derivative
compatible with $f^{(0)}_{ij}$. Note that $f_{(0)}^{ij}F^{(2)}_{ij}=0$
by definition. Thus, we obtain 
%
\begin{equation}
 \tilde{A}_{ij} = \frac{1}{a^3(t)}
  \left[F^{(2)}_{ij}(x^k)\int_{t_0}^t a(t')dt'
  + C^{(2)}_{ij}(x^k)\right] + O(\epsilon^4), 
   \label{eqn: solution-Aij}
\end{equation}
where $C^{(2)}_{ij}(x^k)$ is a symmetric matrix whose components depend
only on the spatial coordinates and which satisfies
$f_{(0)}^{ij}C^{(2)}_{ij}=0$.

Next, the equation (\ref{eqn:eq-for-delta}) for $\delta$ is easily
solved as
%
\begin{equation}
 \delta = \frac{\rho_*a_*^2}{\rho_0(t)a^2(t)}Q^{(2)}(x^k) + O(\epsilon^4),
 \label{eqn: solution-delta}
\end{equation}
where $\rho_*a_*^2$ is a constant and $Q^{(2)}(x^k)$ is an arbitrary
function of the spatial coordinates. With this expression for $\delta$,
(\ref{eqn:eq-for-ui}) gives
%
\begin{equation}
 u_i = \frac{\rho_*a_*^2}{3[\rho_0(t)+P_0(t)]a^3(t)}
  \left[\partial_iQ^{(2)}(x^k)\int_{t_0}^ta(t')dt' +
   C^{(3)}_i(x^k) \right] + O(\epsilon^5),
    \label{eqn:solution-ui}
\end{equation}
where $C^{(3)}_i(x^k)$ is an arbitrary function of the spatial coordinates.

The `constants` of integration $L^{(0)}(x^k)$, $f^{(0)}_{ij}(x^k)$,
$C^{(2)}_{ij}(x^k)$, $Q^{(2)}(x^k)$ and $C^{(3)}_i(x^k)$ are not
independent. Indeed, by solving the Hamiltonian and momentum
constraints (\ref{eqn:Hamiltonian}) and (\ref{eqn:Momentum}), $Q^{(2)}$
and $C^{(3)}_i$ are expressed in terms of other integration `constants` as
%
\begin{eqnarray}
 Q^{(2)} & = & \frac{1}{2\kappa^2\rho_*a_*^2}
  R\left[(L^{(0)})^4f^{(0)}\right] + O(\epsilon^4), 
  \nonumber\\
 C^{(3)}_i & = & \frac{3}{\left(L^{(0)}\right)^6\kappa^2\rho_*a_*^2}
  f_{(0)}^{jk}\tilde{D}^{(0)}_j
  \left[\left(L^{(0)}\right)^6C^{(2)}_{ki}\right] + O(\epsilon^5). 
   \label{eqn: constraint-integration-const}
\end{eqnarray}

Until now, we have not used either (\ref{eqn:p-delta-pi}) or
(\ref{eqn:eq-pi}). Therefore, the general solutions presented above are
valid not only for the scalar field system but also for radiation, dust
or any other sources, provided that the stress-energy tensor is of the
perfect fluid form and that $p=O(\epsilon^2)$.

We now use (\ref{eqn:p-delta-pi}) and (\ref{eqn:eq-pi}) to proceed
further. It is easy to integrate (\ref{eqn:eq-pi}) to give
%
\begin{equation}
 \pi = -\frac{\rho_*a_*^2\partial_t\phi_0}{3(\rho_0+P_0)a^3}
  \left[Q^{(2)}(x^k)\int^t_{t_0}a(t')dt' + \Pi^{(2)}(x^k)\right]
  + O(\epsilon^4), 
\end{equation}
where $\Pi^{(2)}(x^k)$ is an arbitrary function of the spatial
coordinates. By using (\ref{eqn:p-delta-pi}) and (\ref{eqn:evol-K2}) we
obtain 
%
\begin{equation}
 \chi = - \frac{\rho_*a_*^2}{3(\rho_0+P_0)a^2}
  \left[
   \left(1+3c_{s0}^2 - \frac{\rho_0\Gamma_0\partial_t\phi_0}
    {(\rho_0+P_0)a} \int^t_{t_0}a(t')dt'\right) Q^{(2)}(x^k)
   -
   \frac{\rho_0\Gamma_0\partial_t\phi_0}
   {(\rho_0+P_0)a}\Pi^{(2)}(x^k)\right] + O(\epsilon^4). 
  \label{eqn:chi-sol}
\end{equation}
Note that, since
$u_i=-\partial_i\phi/\sqrt{X}=-\partial_i\pi/\partial_t\phi_0+O(\epsilon^5)$, 
the integration `constant` $\Pi^{(2)}(x^k)$ is related to
$C^{(3)}_i(x^k)$ as 
%
\begin{equation}
 C^{(3)}_i = \partial_i\Pi^{(2)} + O(\epsilon^5).
  \label{eqn:constraint-C_3-Pi_2}
\end{equation}
Thus, $C^{(3)}_i$ for the scalar field system does not include a
transverse part.

\subsection{Solution up to $O(\epsilon^2)$}

Solutions obtained so far are correct up to leading order in the
gradient expansion. Among them, the spatial metric $\psi$ and
$\tilde{\gamma}_{ij}$ have been obtained only up to $O(1)$ while all
other variables are correct at least up to $O(\epsilon^2)$. In this
subsection we seek $O(\epsilon^2)$ corrections to $\psi$ and
$\tilde{\gamma}_{ij}$. For this purpose it is convenient to specify the
shift vector $\beta^i$ more accurately than indicated by
(\ref{eqn:order-of-magnitude}): in this subsection we set  
%
\begin{equation}
 \beta = O(\epsilon^3). 
  \label{eqn:additional-gauge-condition}
\end{equation}
With this gauge choice, (\ref{eqn:evol-gamma2}) is reduced to
%
\begin{equation}
  \partial_t \tilde{\gamma}_{ij} = -2\tilde{A}_{ij} + O(\epsilon^4),
\end{equation}
and thus results in 
%
\begin{equation}
 \tilde{\gamma}_{ij} = f^{(0)}_{ij}(x^k) 
  - 2\left[
      F^{(2)}_{ij}(x^k)\int^t_{t_0}\frac{dt'}{a^3(t')}\int^{t'}_{t_0}a(t'')dt''
      + C^{(2)}_{ij}(x^k)\int^t_{t_0}\frac{dt'}{a^3(t')}\right]
  + O(\epsilon^4),
  \label{eqn:gammatilde-sol}
\end{equation}
where we have absorbed a new integration `constant` into the $0$-th
order integration `constant` $f^{(0)}_{ij}(x^k)$. 
Again, we have not used either (\ref{eqn:p-delta-pi}) or
(\ref{eqn:eq-pi}) to derive this result. Therefore, the general
solution (\ref{eqn:gammatilde-sol}) is valid not only for the scalar
field system but also for any other sources, provided that the
stress-energy tensor is of the perfect fluid form, that
$p=O(\epsilon^2)$ and that the additional gauge condition
(\ref{eqn:additional-gauge-condition}) is imposed.

In order to seek the $O(\epsilon^2)$ correction to $\psi$, we specialize
to the scalar field system since we will need the solution
(\ref{eqn:chi-sol}) for $\chi$, which was obtained by using 
(\ref{eqn:p-delta-pi}). The evolution equation (\ref{eqn:evol-gamma1})
with (\ref{eqn:psi-leading}) leads to 
%
\begin{equation}
 \ln\left[\frac{\psi}{L^{(0)}(x^k)}\right] =
  \frac{1}{2}\int^t_{t_0}H(t')\chi(t')dt' + O(\epsilon^4),
  \label{eqn: solution-psi}
\end{equation}
where we have absorbed a new integration `constant` into the $0$-th
order integration `constant` $L^{(0)}(x^k)$. We can substitute
(\ref{eqn:chi-sol}) to the right hand side of this equation to complete
the procedure.

\subsection{Summary of the solution}
\label{subsec:summary-solution}

In summary we have obtained the following solutions in the gradient
expansion for the scalar field system. 
%
\begin{eqnarray}
 \delta & = & \frac{1}{2\kappa^2\rho_0a^2}
  R\left[(L^{(0)})^4f^{(0)}\right]  + O(\epsilon^4), \nonumber\\
 u_i & = & \frac{1}{6\kappa^2(\rho_0+P_0)a^3}
  \partial_i 
  \left(R\left[(L^{(0)})^4f^{(0)}\right] \int^t_{t_0}a(t')dt' +
   C^{(2)}\right) + O(\epsilon^5), \nonumber\\
 \pi & = & -\frac{\partial_t\phi_0}{6\kappa^2(\rho_0+P_0)a^3}
  \left(R\left[(L^{(0)})^4f^{(0)}\right] \int^t_{t_0}a(t')dt' +
   C^{(2)}\right)
  + O(\epsilon^4),  \nonumber\\
 \chi & = & - \frac{1}{6\kappa^2(\rho_0+P_0)a^2}
  \left[
   \left(1+3c_{s0}^2 - \frac{\rho_0\Gamma_0\partial_t\phi_0}
    {(\rho_0+P_0)a} \int^t_{t_0}a(t')dt'\right) 
   R\left[(L^{(0)})^4f^{(0)}\right]
   -
   \frac{\rho_0\Gamma_0\partial_t\phi_0}
   {(\rho_0+P_0)a}C^{(2)}\right]\nonumber\\
 & & + O(\epsilon^4), \nonumber\\
 \psi & = & L^{(0)}
  \left(1+
  \frac{1}{2}\int^t_{t_0}H(t')\chi(t')dt'\right) + O(\epsilon^4),  
  \nonumber\\
 \tilde{\gamma}_{ij} & = & f^{(0)}_{ij}
  - 2\left(
      F^{(2)}_{ij}\int^t_{t_0}\frac{dt'}{a^3(t')}\int^{t'}_{t_0}a(t'')dt''
      + C^{(2)}_{ij}\int^t_{t_0}\frac{dt'}{a^3(t')}\right)
  + O(\epsilon^4), \nonumber\\
 \tilde{A}_{ij} & = & \frac{1}{a^3}
  \left(F^{(2)}_{ij}\int_{t_0}^t a(t')dt' + C^{(2)}_{ij}\right) 
  + O(\epsilon^4),
  \label{eqn:general solution}
\end{eqnarray}
where $C^{(2)}$ in this expression  is related to $\Pi^{(2)}$ in 
(\ref{eqn:chi-sol})  as $C^{(2)}=2\kappa^2\rho_* a_*^2 \Pi^{(2)}$, and 
$F^{(2)}_{ij}$ is defined by (\ref{eqn:def-F2ij}), and `constants`
of integration $L^{(0)}$, $f^{(0)}_{ij}$, $C^{(2)}$ and $C^{(2)}_{ij}$
depend only on the spatial coordinates $\{x^k\}$ ($k=1,2,3$), and satisfy 
%
\begin{eqnarray}
 f^{(0)}_{ij} & = & f^{(0)}_{ji}, \quad \det(f^{(0)}_{ij}) = 1,
  \nonumber\\
 C^{(2)}_{ij} & = & C^{(2)}_{ji}, \quad f_{(0)}^{ij}C^{(2)}_{ij} = 0,
  \nonumber\\
 \left(L^{(0)}\right)^6 \partial_iC^{(2)} & = & 
  6 f_{(0)}^{jk}\tilde{D}^{(0)}_j
  \left[\left(L^{(0)}\right)^6C^{(2)}_{ki}\right]. 
\end{eqnarray}
Here, $f_{(0)}^{ij}$ is the inverse matrix of $f^{(0)}_{ij}$ and
$\tilde{D}^{(0)}$ is the covariant derivative compatible with
$f^{(0)}_{ij}$. 

Note that the gauge condition (\ref{eqn:additional-gauge-condition}) is
unchanged under purely spatial coordinate transformations
%
\begin{equation}
 x^i \to {x'}^i = f^i(x^k).
\end{equation}
Thus, the $0$-th order spatial metric $f^{(0)}_{ij}$ includes $3$ gauge
degrees of freedom. Therefore, the number of degrees of freedom included
in each `constant` of integration is 
%
\begin{eqnarray}
 L^{(0)} & \cdots & 
  1 \mbox{ scalar growing mode }
  = 1 \mbox{ component }, \nonumber\\
 f^{(0)}_{ij} & \cdots & 
  2 \mbox{ tensor growing modes }
  = 5 \mbox{ components } - 3 \mbox{ gauge }, \nonumber\\
 C^{(2)} & \cdots & 
  1 \mbox{ scalar decaying mode }
  = 1 \mbox{ component }, \nonumber\\
 C^{(2)}_{ij} & \cdots & 
  2 \mbox{ tensor decaying modes }
  = 5 \mbox{ components } - 3 \mbox{ constraints }. 
\end{eqnarray}

\section{Examples}
\label{sec:examples}

Until now, we have not specified a form of the function $P(X,\phi)$. In
this section, we shall consider some specific examples. 

\subsection{Scalar with shift symmetry}
\label{subsec:PX}

Let us consider the case where the Lagrangian $P$ depends on the scalar
field only through $X$, i.e. $P=P(X)$. This case corresponds to a scalar
with shift symmetry and is often considered in models of
k-inflation~\cite{ArmendarizPicon:1999rj}.

The sound speed $c_{s0}^2$ and $\Gamma_0$ in this case are 
\begin{align}
c_{s0}^2={P_{0X}\over \rho_{0X}}\,,\quad 
\Gamma_0=0\,,
\end{align}
and the sound speed agrees with the adiabatic sound speed defined by
$\delta P=c_s^2\delta\rho$. From this fact, it is expected that
dynamics of the scalar field system should somehow resemble that of a
perfect fluid with equation of state $P=P(\rho)$. This expectation turns
out to be essentially correct. In fact, in Appendix we see that these
two systems have essentially the same general solutions up to
$O(\epsilon^2)$ in the gradient  expansion except for the following one
difference. The $3$-velocity $u_i$ for the scalar field system does not
include transverse mode but that for the perfect fluid in general
does. The evolution of other quantities such as $\delta$, $\chi$,
$\psi$, $\tilde{\gamma}_{ij}$ and $\tilde{A}_{ij}$ are the same.

Note that ghost
condensation~\cite{ArkaniHamed:2003uy,ArkaniHamed:2005gu,Mukohyama:2006be}
also has a similar low energy effective Lagrangian but includes terms
like $(\vec{\nabla}^2\pi)^2$ as well, where $\vec{\nabla}$ is the
spatial gradient  and $\pi$ is perturbation of the scalar field. Indeed,
those additional terms play important roles in generation of primordial
density perturbations~\cite{ArkaniHamed:2003uz} and infra-red
modification of gravity~\cite{ArkaniHamed:2003uy}. In other words, as
easily seen by doing a proper scaling analysis, a Lagrangian of the form
$P(X)$ {\it without} those additional terms can {\it not} describe ghost 
condensation in general even at low energies. However, if we are
interested in a situation where the the Hubble scale during inflation is
longer than the scale of IR modification of gravity then we can safely
use the present formalism to investigate classical evolution of
cosmological perturbation at superhorizon scales. However, if the scale
of IR modification is longer then the present formulation is not
valid. In this case we probably need to extend the present formulation
to include the higher derivative term. More detailed investigation will
be considered in future publication.

\subsection{Canonical scalar}
\label{subsec:canonical}

Next let us consider a canonical scalar field, i.e. 
\begin{equation}
P(X,\phi)={X\over 2}-V(\phi)\,,
\end{equation}
where $V(\phi)$ is a potential. In this case, we can obtain 
\begin{align}
c_{s0}^2=1\,,\quad \Gamma_0=-{2V_{\phi0}\over \rho_0}\,,
\end{align}
where $V_{\phi0}\equiv dV(\phi_0)/d\phi_0$. The general solution can be
read off from (\ref{eqn:general solution}). In particular, $\chi$ is
given by 
\begin{align}
\chi = -\frac{2\rho_*a_*^2}{3\dot{\phi}_0^3 a^3}\left[ 
  \left(2a\phi_0+V_{\phi0}\int^t_{t_0}a(t')dt'\right)Q^{(2)}(x^k)+
    V_{\phi0}\Pi^{(2)}(x^k)\right] + O(\epsilon^4), 
    \label{eqn:chi-sol-canonicalscalar}
\end{align}
where $\dot{\phi}_0=\partial_t{\phi}_0$. If we set 
$Q^{(2)}(x^k) \rho_* a_*^2=-3\dot{\phi}_* C(x^k)$ and 
$\Pi^{(2)}(x^k) \rho_* a_*^2=-3\dot{\phi}_* D(x^k)$, 
then this reduces to eq.~(C29) of \cite{Tanaka:2007gh}.

\subsection{DBI scalar}

We now consider a scalar field described by the so called DBI
action. This kind of scalar is considered as an inflaton in an
interesting class of inflationary models called DBI
inflation~\cite{Silverstein:2003hf,Alishahiha:2004eh}. In a simple case
the inflaton scalar field corresponds to the radial position of a
D$3$-brane in a warped compactification.

For a warped throat background, the 10-dimensional metric takes the
following form 
\begin{equation}
ds^2=h^2(\rho)\eta_{\mu\nu}dx^\mu dx^\nu+h^{-2}(\rho)
\left(d\rho^2+\rho^2 g_{mn}^{(5)}dx^m dx^n \right )\,,
\end{equation}
where $h(\rho)$ is a warp factor, $x^\mu ~(\mu=0,1,2,3)$ are external
$4$-dimensional coordinates, $\rho$ is the radial coordinate in the
warped throat, and $x^m~(m=5,6,7,8,9)$ are the internal $5$-dimensional
angular coordinates. Considering the radial position $\rho$ of a
D$3$-brane in this background as a scalar field in the external
$4$-dimensional spacetime, its dynamics is described by a
Dirac-Born-Infeld (DBI) action plus a Chern-Simons term and additional
potential terms. Thus, the Lagrangian $P(X,\phi)$ is 
\begin{equation}
 P(X,\phi)=-T(\phi)\sqrt{1-X/T(\phi)}+T(\phi)-V(\phi)\,,
  \label{eqn:P-DBI}
\end{equation}
where $\phi\equiv T_3^{1/2}\rho$ is the scalar field describing the
D$3$-brane position, $T(\phi)\equiv T_3 h^4(\phi)$ is the warped brane
tension, and $V(\phi)$ is the inflaton potential. Here, $T_3$ is the
D$3$-brane tension. The Lagrangian (\ref{eqn:P-DBI}) also applies to the
case where the inflaton $\phi$ is the radial position of a wrapped D$5$-
or D$7$-brane~\cite{Kobayashi:2007hm}.

For the Lagrangian (\ref{eqn:P-DBI}) we obtain
\begin{align}
c_{s0}^2={1\over \gamma^2}\,,\quad 
\Gamma_0={1\over \rho_0}
\left[-{2T_{\phi0}\over \gamma}+(T_{\phi0}-V_{\phi0})
\left(1+{1\over \gamma^2}\right)\right], 
\end{align}
where $T_{\phi0}\equiv dT(\phi_0)/d\phi_0$ and 
$\gamma\equiv 1/\sqrt{1-X_0/T(\phi_0)}$. In a non-relativistic limit
($\gamma\simeq 1$), $c_{s0}^2$ and $\Gamma_0$ reduce to those for the
canonical scalar field discussed in subsection~\ref{subsec:canonical}.

\section{Summary and discussion}
\label{sec:summary}

We have developed a theory of nonlinear cosmological perturbations on
superhorizon scales for a scalar field described by a Lagrangian of the
form $P(X,\phi)$, where $X=-\partial^{\mu}\phi\partial_{\mu}\phi$  and
$\phi$ is the scalar field, and also for a perfect fluid with a general
equation of state $P=P(\rho)$. The general solutions valid up to the
order $O(\epsilon^2)$ in the spatial gradient expansion have been
presented in subsection~\ref{subsec:summary-solution} for the scalar
field system and in Appendix~\ref{app:perfectfluid} for the perfect
fluid. 

This formalism can be applied to many interesting circumstances. Some
particular examples have been listed in Sec.~\ref{sec:examples},
including a scalar with shift symmetry, a canonical scalar and a DBI
scalar. (As shown in Appendix~\ref{app:perfectfluid} it can be applied 
also for a perfect fluid with general equation of state
$P=P(\rho)$.) Thus, the formalism can be used to investigate
superhorizon evolution of nonlinear cosmological perturbations in
k-inflation and DBI inflation. As explained in Sec.~\ref{sec:intro},
non-Gaussianity can be affected by order $O(\epsilon^2)$ corrections if,
e.g. there is a stage which violates the slow roll conditions.

For ghost inflation, applicability of the present formulation seems
a bit subtle as briefly discussed in subsection~\ref{subsec:PX}. 
If the Hubble scale is longer than the scale of IR modification of
gravity then we can use the present formalism to investigate
superhorizon perturbations. On the other hand, if one is interested in
a situation where the scale of IR modification is longer then the
present formulation is not valid. We hope to address this issue in more 
detail in future publication.

Recently, models of multi-field DBI have been also studied to 
investigate large non-Gaussianity by \cite{Arroja:2008yy,Langlois:2008qf}.  
Here our formulation has been developed in a single scalar field, however, 
we also plan to extend it to a system of multi-field scalar in the future.

\acknowledgments
Y.T. would like to thank Kei-ichi Maeda, Tsutomu
Kobayashi, and Shuichiro Yokoyama for their comments and/or discussions
on this work, and also wish to show our acknowledgement to the 
financial support by Waseda University. 
The work of S.M. was supported in part by MEXT through a Grant-in-Aid
for Young Scientists (B) No.~17740134, and by JSPS through a
Grant-in-Aid for Creative Scientific Research No.~19GS0219 and through a
Grant-in-Aid for Scientific Research (B) No.~19340054. This work was
supported by World Premier International Research Center Initiative¡ÊWPI
Initiative), MEXT, Japan.

\appendix
\section{Perfect fluid with $P=P(\rho)$}
\label{app:perfectfluid}

In this appendix, we shall consider a universe dominated by a perfect
fluid with a general equation of state $P=P(\rho)$. The stress-energy
tensor is given by 
\begin{equation}
 T_{\mu\nu}=(\rho+P)u_\mu u_\nu+Pg_{\mu\nu}\,,\quad P=P(\rho)\,.
\end{equation}
The formulation in this case is similar to that presented in
Sec.~\ref{sec:formulation} and Sec.~\ref{sec:solution} for a scalar 
field. Main differences come from the form of the pressure perturbation
$p$.

As in Sec.~\ref{sec:formulation} and Sec.~\ref{sec:solution}, we employ
the ADM formalism and the gradient expansion in the uniform Hubble
slicing. In the ADM formalism, the dynamical equations are given by
(\ref{eqn:dynamica-eqs}), with the two constraint equations 
(\ref{eqn:Hamiltonian-const}) and (\ref{eqn:Momentum-const}). We further 
decompose the spatial metric and the extrinsic  curvature as
(\ref{eqn:decompose-metric}).  Then we obtain the first-order equations
(\ref{eqn:dpsi}) and (\ref{eqn:dgammatilde}) for the spatial metric 
$(\psi, \til{\gamma}_{ij})$,  
and (\ref{eqn:dynam-K-K}) and (\ref{eqn:dynam-K-tildeA}) 
for the extrinsic curvature $(K, \til{A}_{ij})$. The Hamiltonian and
momentum constraint are, respectively,
(\ref{eqn:Hamiltonian-constraint})  and
(\ref{eqn:momentum-constraint}). The conservation equation 
$\nabla_\mu T^\mu_\nu=0$ is given by (\ref{eqn:conservation-t}) 
and (\ref{eqn:conservation-i}). We
adopt the uniform Hubble slicing  as (\ref{eqn:uniform-Hubble}).

Next, we will employ the gradient expansion. In this approach 
we introduce a flat FRW universe
($a(t)$, $\rho_0(t)$) as a background  and suppose that the 
characteristic length scale $L$ of perturbations is longer than the
Hubble length scale $1/H$ of the background, i.e.  $HL\gg 1$. Therefore,
we consider $\epsilon\equiv 1/(HL)$ as a small parameter and
systematically expand our equations by $\epsilon$, considering a spatial
derivative acted on perturbations is of order $O(\epsilon)$. 
The background flat FRW universe ($a(t)$, $\rho_0(t)$) satisfies the
Friedmann equation and the conservation equation
$\partial_t\rho_0+3H(\rho_0+P_0)=0$, where $P_0=P(\rho_0)$. 
We can estimate order of magnitude of various quantities by using the
assumption (\ref{eqn:assumption-vbeta}) and
(\ref{eqn:assumption-gamma}). As a result, we have the estimates shown
in (\ref{eqn:order-of-magnitude}) except that the condition
$\pi=O(\epsilon^2)$ does not exist in the present case. Then
substituting the order of magnitude (\ref{eqn:order-of-magnitude}) into
the basic equations gives leading order equations as
(\ref{eqn:hydro1})-(\ref{eqn:evol-K2}).

By using these equations and background conservation equation, equations
for $\delta$ and $u_i$ are easily obtained as (\ref{eqn:eq-for-delta})
and (\ref{eqn:eq-for-ui}). They are easily solved as 
(\ref{eqn: solution-delta}) and (\ref{eqn:solution-ui}). The traceless
part of the extrinsic curvature $\til{A}_{ij}$ is solved by using the
leading part of $\psi$ and $\til{\gamma}_{ij}$, (\ref{eqn:psi-leading})
and (\ref{eqn:gammatilde-leading}), as (\ref{eqn: solution-Aij}). The
`constants` of integration are not independent but are related to each
other by the two constraint equations as 
(\ref{eqn: constraint-integration-const}). However, in the present case 
there is no relation like (\ref{eqn:constraint-C_3-Pi_2}), and then
$C^{(3)}_i$ contains transverse component as well as the longitudinal
component. 

For $P=P(\rho)$, perturbation of the pressure $p\equiv P-P_0$ is given
by 
\begin{align}
p=\rho_0 c_{s0}^2 \delta +O(\epsilon^4)\,,\quad
 c_{s0}^2=\frac{dP_0}{d\rho_0}. 
\end{align}
By using (\ref{eqn:evol-K2}), we obtain the solution for $\chi$ as 
\begin{align}
\chi  =  - \frac{(1+3c_{s0}^2)R\left[(L^{(0)})^4f^{(0)}\right]}
{6\kappa^2(\rho_0+P_0)a^2}+O(\epsilon^4)\,.
\label{eqn: solution-chi}
\end{align}
In order to obtain a general solution valid up to $O(\epsilon^2)$, 
we further seek $O(\epsilon^2)$ corrections to $\psi$ and
$\tilde{\gamma}_{ij}$. For this purpose we adopt the gauge condition
(\ref{eqn:additional-gauge-condition}). With this gauge choice, the
$O(\epsilon^2)$ part of $\til{\gamma}_{ij}$ can be obtained by  
using the solution of $\til{A}_{ij}$ as (\ref{eqn:gammatilde-sol}). 
Similarly, by using the solution of $\chi$ (\ref{eqn: solution-chi}), 
the $O(\epsilon^2)$ correction to $\psi$ is obtained as 
(\ref{eqn: solution-psi}), provided that $\chi$ in the present case is
given by (\ref{eqn: solution-chi}). 

In summary we have obtained a general solution valid up to
$O(\epsilon^2)$ for the perfect fluid with $P=P(\rho)$. 
\begin{eqnarray}
 \delta & = & \frac{1}{2\kappa^2\rho_0a^2}
  R\left[(L^{(0)})^4f^{(0)}\right]  + O(\epsilon^4), \nonumber\\
 u_i & = & \frac{1}{6\kappa^2(\rho_0+P_0)a^3}
  \left(\partial_i R\left[(L^{(0)})^4f^{(0)}\right] \int^t_{t_0}a(t')dt' +
   C^{(3)}_i\right) + O(\epsilon^5), \nonumber\\
 \chi & = &  - \frac{1+3c_{s0}^2}{6\kappa^2(\rho_0+P_0)a^2}
  R\left[(L^{(0)})^4f^{(0)}\right]
  +O(\epsilon^4), \nonumber\\
 \psi & = & L^{(0)}
  \left(1+
  \frac{1}{2}\int^t_{t_0}H(t')\chi(t')dt'\right) + O(\epsilon^4),  
  \nonumber\\
 \tilde{\gamma}_{ij} & = & f^{(0)}_{ij}
  - 2\left(
      F^{(2)}_{ij}\int^t_{t_0}\frac{dt'}{a^3(t')}\int^{t'}_{t_0}a(t'')dt''
      + C^{(2)}_{ij}\int^t_{t_0}\frac{dt'}{a^3(t')}\right)
  + O(\epsilon^4), \nonumber\\
 \tilde{A}_{ij} & = & \frac{1}{a^3}
  \left(F^{(2)}_{ij}\int_{t_0}^t a(t')dt' + C^{(2)}_{ij}\right) 
  + O(\epsilon^4),
  \label{eqn:general solution-perfectfluid}
\end{eqnarray}
where 
\begin{align}
 C^{(3)}_i =  \frac{3}{\left(L^{(0)}\right)^6\kappa^2\rho_*a_*^2}
  f_{(0)}^{jk}\tilde{D}^{(0)}_j
  \left[\left(L^{(0)}\right)^6C^{(2)}_{ki}\right] + O(\epsilon^5). 
\end{align}
Here $F^{(2)}_{ij}$ is defined by (\ref{eqn:def-F2ij}), and `constants`
of integration $L^{(0)}$, $f^{(0)}_{ij}$ and $C^{(2)}_{ij}$
depend only on the spatial coordinates $\{x^k\}$ ($k=1,2,3$) and satisfy 
%
\begin{eqnarray}
 f^{(0)}_{ij} & = & f^{(0)}_{ji}, \quad \det(f^{(0)}_{ij}) = 1,
  \nonumber\\
 C^{(2)}_{ij} & = & C^{(2)}_{ji}, \quad f_{(0)}^{ij}C^{(2)}_{ij} = 0. 
\end{eqnarray}
Here, $f_{(0)}^{ij}$ is the inverse matrix of $f^{(0)}_{ij}$ and
$\tilde{D}^{(0)}$ is the covariant derivative compatible with
$f^{(0)}_{ij}$. Compared with the solution (\ref{eqn:general solution})
for the scalar field system, only differences are: (i) $\pi$ does not
exist; (ii) $\Gamma_0=0$; and (iii) $C^{(3)}_i$ includes not only a
longitudinal component but also transverse components.

The number of degrees of freedom included in each `constant` of integration is 
%
\begin{eqnarray}
 L^{(0)} & \cdots & 
  1 \mbox{ growing adiabatic mode of density perturbation}
  = 1 \mbox{ component }, \nonumber\\
 f^{(0)}_{ij} & \cdots & 
  2 \mbox{ tensor growing modes }
  = 5 \mbox{ components } - 3 \mbox{ gauge }, \nonumber\\
 C^{(2)}_{ij} & \cdots & 
  2 \mbox{ tensor decaying modes }+ 
  3 \mbox{ velocity }
  = 5 \mbox{ components }. 
\end{eqnarray}

\end{document}